\documentclass[12pt]{article}
\usepackage{epsfig}

\newcommand\pubnumber{CERN-TH/2001-003}

\newcommand\hepnumber{hep-ph/0101121}

\def\Title#1{\begin{center} {\Large\bf #1 } \end{center}}
\def\Author#1{\begin{center}{ \sc #1} \end{center}}
\def\Address#1{\begin{center}{ \it #1} \end{center}}

\newcommand\pubblock{\rightline{\begin{tabular}{l} \pubnumber\\
          \hepnumber \end{tabular}}}
\newenvironment{Abstract}{\begin{quotation}  }{\end{quotation}}
\newenvironment{Presented}{\begin{quotation} \begin{center} 
             Presented at the\end{center}
      \begin{center}\begin{large}}{\end{large}\end{center} \end{quotation}}

\makeatletter
\def\section{\@startsection{section}{0}{\z@}{5.5ex plus .5ex minus
 1.5ex}{2.3ex plus .2ex}{\large\bf}}
\def\subsection{\@startsection{subsection}{1}{\z@}{3.5ex plus .5ex minus
 1.5ex}{1.3ex plus .2ex}{\normalsize\bf}}
\def\subsubsection{\@startsection{subsubsection}{2}{\z@}{-3.5ex plus
-1ex minus  -.2ex}{2.3ex plus .2ex}{\normalsize\sl}}

\renewcommand{\@makecaption}[2]{%
   \vskip 10pt
   \setbox\@tempboxa\hbox{\small #1: #2}
   \ifdim \wd\@tempboxa >\hsize     
       \small #1: #2\par          
     \else                        
       \hbox to\hsize{\hfil\box\@tempboxa\hfil}
   \fi}

 \def\citenum#1{{\def\@cite##1##2{##1}\cite{#1}}}
 
\newcount\@tempcntc
\def\@citex[#1]#2{\if@filesw\immediate\write\@auxout{\string\citation{#2}}\fi
  \@tempcnta\z@\@tempcntb\m@ne\def\@citea{}\@cite{\@for\@citeb:=#2\do
    {\@ifundefined
       {b@\@citeb}{\@citeo\@tempcntb\m@ne\@citea\def\@citea{,}{\bf ?}\@warning
       {Citation `\@citeb' on page \thepage \space undefined}}%
    {\setbox\z@\hbox{\global\@tempcntc0\csname b@\@citeb\endcsname\relax}%
     \ifnum\@tempcntc=\z@ \@citeo\@tempcntb\m@ne
       \@citea\def\@citea{,}\hbox{\csname b@\@citeb\endcsname}%
     \else
      \advance\@tempcntb\@ne
      \ifnum\@tempcntb=\@tempcntc
      \else\advance\@tempcntb\m@ne\@citeo
      \@tempcnta\@tempcntc\@tempcntb\@tempcntc\fi\fi}}\@citeo}{#1}}
\def\@citeo{\ifnum\@tempcnta>\@tempcntb\else\@citea\def\@citea{,}%
  \ifnum\@tempcnta=\@tempcntb\the\@tempcnta\else
  {\advance\@tempcnta\@ne\ifnum\@tempcnta=\@tempcntb \else\def\@citea{--}\fi
    \advance\@tempcnta\m@ne\the\@tempcnta\@citea\the\@tempcntb}\fi\fi}
\makeatother

\newcommand{\beq}{\begin{equation}}
\newcommand{\eeq}{\end{equation}}
\newcommand{\ba}{\begin{array}}
\newcommand{\ea}{\end{array}} 
\newcommand{\beqa}{\begin{eqnarray}}
\newcommand{\eeqa}{\end{eqnarray}}

\newcommand{\cL}{{\cal L}}

\newcommand{\cA}{{\cal A}}
\newcommand{\cO}{{\cal O}}

\newcommand{\Br}{{\cal B}}

\newcommand{\no}{\nonumber}
\newcommand{\lsim}{\stackrel{<}{_\sim}}
\newcommand{\gsim}{\stackrel{>}{_\sim}}

\newcommand{\op}{Q}

\newcommand{\cne}{C_9^{\rm eff}}
\renewcommand{\Im}{{\rm Im}}

\newcommand{\ct}{C_{10}}
\newcommand{\klpn}{K_L \to \pi^0 \nu \bar \nu}
\newcommand{\kppn}{K^+ \to \pi^+ \nu \bar \nu}

\newcommand{\kpe}{K_L \to \pi^0 e^+ e^-}

\def\npb#1#2#3{    {\it Nucl. Phys. }{\bf B#1} (#2) #3}
\def\plb#1#2#3{    {\it Phys. Lett. }{\bf B#1} (#2) #3}
\def\prd#1#2#3{    {\it Phys. Rev. }{\bf D#1} (#2) #3}

\def\prl#1#2#3{    {\it Phys. Rev. Lett. }{\bf #1} (#2) #3}
\def\ptp#1#2#3{    {\it Prog. Theor. Phys. }{\bf #1} (#2) #3}

\def\rmp#1#2#3{    {\it Rev. Mod. Phys. }{\bf #1} (#2) #3}
\def\zpc#1#2#3{    {\it Z. Phys. }{\bf C#1} (#2) #3}

\def\jhep#1#2#3{   {\it JHEP  }{\bf #1} (#2) #3}
\def\nobook#1#2#3{ {\bf #1} (#2) #3}

\begin{document}
\begin{titlepage}
\pubblock

\vfill
\def\thefootnote{\fnsymbol{footnote}}
\Title{Supersymmetric effects in rare \\[5pt]  semileptonic decays of $B$ 
and $K$ mesons}
\vfill
\Author{Gino Isidori}
\Address{Theory Division, CERN, CH-1211 Geneva 23, 
Switzerland~\footnote{On leave from INFN, Laboratori Nazionali di 
Frascati, Via Enrico Fermi 40, I-00044 Frascati (Rome), Italy.}}
\vfill
\begin{Abstract}
Rare flavour-changing neutral-current transitions of the type  
$s \to d~\ell^+\ell^-(\nu\bar\nu)$ and 
$b \to s~\ell^+\ell^-(\nu\bar\nu)$
are analysed in supersymmetric extensions
of the Standard Model with generic flavour couplings. 
It is shown that these processes are particularly sensitive 
to the left--right mixing of the squarks and that,
in the presence of non-universal $A$ terms, they
could lead to unambiguous signatures of 
new physics in exclusive $K$ and $B$ meson decays.
\end{Abstract}
\vfill
\begin{Presented}
5th International Symposium on Radiative Corrections \\ 
(RADCOR--2000) \\[4pt]
Carmel CA, USA, 11--15 September 2000
\end{Presented}
\vfill
\end{titlepage}
\def\thefootnote{\arabic{footnote}}
\setcounter{footnote}{0}

\section{Introduction}
Flavour-changing neutral-current (FCNC) processes are one of the most 
powerful tools in probing the structure of flavour beyond the   
Standard Model (SM): the strong suppression of these 
transitions occurring within the SM, which is 
due to the Glashow--Iliopoulos--Maiani (GIM) mechanism \cite{GIM} and to
the hierarchy of the Cabibbo--Kobayashi--Maskawa (CKM) matrix \cite{CKM},
ensures a large sensitivity to possible non-standard effects, 
even if these occur at very high energy scales.

In the present talk we focus on a specific class of 
$\Delta F=1$ FCNC transitions:
\beq
s \to d~\ell^+\ell^-(\nu\bar\nu) \qquad {\rm and}\qquad  
b \to s~\ell^+\ell^-(\nu\bar\nu)~.
\label{eq:uno}
\eeq
As we shall discuss, these are particularly interesting 
for the following reasons:
\begin{itemize}
\item{} 
These transitions have a strong sensitivity to supersymmetric extensions 
of the SM with flavour non-universal soft-breaking terms. 
Taking into account all the 
existing phenomenological constraints, within this type of models it is 
possible to generate sizeable non-standard effects to the partonic 
processes in (\ref{eq:uno}).
\item{} The existing experimental constraints on these transitions are rather 
weak, but in the near future it will be possible to perform stringent 
tests by means of exclusive rare $K$ and $B$ meson decays.
\end{itemize}

In the rest of the talk we shall illustrate 
these two points in more detail. 
Section~2 is devoted to the analysis of the 
supersymmetric contributions  to $d_j \to d_i \ell^+\ell^-(\nu\bar\nu)$ 
transitions, including a discussion about the indirect bounds 
obtained by other processes.
In Sections 3--5 we analyse how to extract information on these 
partonic transitions by means of experimental data on 
$K \to \pi \nu\bar{\nu}$, $K_L \to \pi^0 e^+ e^-$ and exclusive 
$b\to s \ell^+ \ell^-(\nu\bar\nu)$ decays, respectively.

\section{SUSY contributions to $d_j \to d_i \ell^+\ell^-(\nu\bar\nu)$ 
transitions in models with non-universal soft-breaking terms}

The class of supersymmetric extensions of the SM that 
we shall consider is the so-called unconstrained MSSM 
(see e.g. \cite{HKR,GGMS}).
This model has the minimal number of new fields necessary 
to build a consistent SUSY version of the SM, namely
all the superpartners of the SM fields plus an extra 
SUSY Higgs doublet. On the contrary, the assumptions made on 
the soft-breaking terms are very general. The only condition 
we shall impose on the flavour structure of the soft-breaking terms 
is a non-universal linear relation between the trilinear terms ($Y^A_{ij}$) 
and the Yukawa couplings ($y_k$), leading to 
\beq
Y^A_{ij}=\cO( y_k M_S)~, \qquad\qquad k={\rm max}(i,j)~,
\label{CDcond}
\eeq
where $M_S$ denotes a common soft-breaking scale 
[$M_{ij}=\cO(M_S^2)$ for the bilinear terms].
This condition let us to avoid charge- and colour-breaking 
minima or unbounded directions in the SUSY potential \cite{CD}. 
The proportionality coefficients will be assumed to be $\cO(1)$, 
unless more stringent constraints are imposed by experimental data.

Similarly to the SM, also within this context FCNC 
amplitudes involving external quark fields turn out 
to be generated only at the quantum level.
Given the large number of new off-diagonal flavour 
couplings, the simplest way to parametrize the new 
effect is provided by the so-called mass-insertion approximation
\cite{HKR,GGMS}. This consists of choosing a simple basis for the gauge 
interactions and, in that basis, to perform a 
perturbative expansion of the squark mass matrices
around their diagonal. Being interested in processes 
with external down-type quarks,  we will employ 
in the following 
a squark basis where all quark--squark--gaugino 
vertices involving down-type quarks are flavour-diagonal.
In this basis we then define the following adimensional couplings:
\beq
\left(\delta^{[U,D]}_{AB}\right)_{ij}=\left(M^2_{[U,D]}\right)_{i_A j_B}
\left/ \langle M^2_{[U,D]} \rangle \right.~,
\label{deltas}
\eeq
where $A,B$ denote the helicity ($L,R$) and $i,j$ the family.
These couplings constitute the basic tool to parametrize and classify  
the new contributions to FCNC amplitudes arising within the 
unconstrained MSSM.

SUSY contributions to  $d_j \to d_i \ell^+\ell^-(\nu\bar\nu)$
transitions can also be divided into three groups according 
to the diagrams (or the effective operators) that generate it:
box and helicity-conserving photon-penguins (generic dimension-6 operators), 
magnetic penguins (dimension-5 operators) and $Z$ penguins. 
In each of these classes a different type of delta plays a
dominant role.

\begin{description}
\item[Generic dimension-6 operators.]
Box diagrams with internal chargino or neutralino 
fields and, in the case of charged leptons, also photon-penguin 
diagrams with internal gluino, chargino or neutralino fields,
can lead to effective FCNC operators of the type 
\beq
\left( {\bar d}_A^i \gamma^\mu d_A^j \right)
\left( {\bar l}_B \gamma_\mu l_B \right)~.
\label{eq:dim6}
\eeq
Since the external quarks have the same helicity, the potentially 
leading SUSY contributions to the Wilson coefficients of these 
operators are generated by helicity-conserving couplings:
\beq
\frac{ \left(\delta^{[U,D]}_{AA}\right)_{ij}  }{ M_S^2}~.
\label{eq:LLmi}
\eeq 
The dimensional factor in Eq.~(\ref{eq:LLmi}), due to
the integration of heavy SUSY degrees of freedom, indicates 
explicitly that these contributions vanish as $1/M^2_S$ 
in the limit of a large SUSY-breaking scale.

The helicity-violating couplings $\delta^Q_{LR}$ appear in 
the Wilson coefficients of dimen\-sion-6 operators
only to second order in the mass expansion, with 
contributions of the type~\cite{CI}
\beq
\frac{ (\delta^U_{LR})_{i3}  (\delta^U_{RL})_{3j} }{ M_S^2}~.
\label{eq:2mi}
\eeq
Since the left--right mixing is generated by the
trilinear terms, then $\delta^Q_{LR}= \cO(m_q /M_S)$ and 
the contribution in (\ref{eq:2mi}) vanish as $1/M^4_S$ for large $M_S$.
Thus the effect of helicity-violating couplings
is not only disfavoured by the fact that it requires a double 
insertion, but it is also parametrically suppressed in the 
limit of a large SUSY-breaking scale. As we shall see below, 
this is not the case only in a specific type 
of dimension-6 operators: those generated by $Z$ penguins.

On the other hand, both helicity-conserving and 
helicity-violating contributions to generic dimension-6 operators
turn out to be negligible with 
respect to the SM ones, once the bounds from 
$\Delta F=2$ processes are taken into account \cite{BRS}. 
This fact can be understood by a naive dimensional 
argument in the limit of large $M_S$ \cite{BCIRS}. 
Indeed, considering for simplicity only the case of 
$\delta^Q_{AA}$,  it is easy to show that the SUSY contribution to  
$\Delta F=2$ amplitudes --appearing necessarily at the 
second order in the mass expansion--
are of  $\cO[(\delta^Q_{AA}/M_S)^2]$. Thus the 
limits on $\delta^Q_{AA}$ arising from 
$\Delta F=2$ amplitudes scale 
linearly with $M_S$ and not quadratically, as
in the $\Delta F=1$ case.
As a result, SUSY contributions to 
$d_j \to d_i \ell^+\ell^-(\nu\bar\nu)$ transitions 
generated by box diagrams and helicity-conserving 
photon-penguins turn out to be 
extremely suppressed for  
$M_S\gsim 1$~TeV.\footnote{A similar argument holds for 
SUSY contributions to $d_j \to d_i \bar q q$
transitions \cite{BCIRS}, with the notable exception 
of $\Delta I=3/2$ amplitudes \protect\cite{KN}.}

\item[Magnetic penguins.]
The integration of the heavy SUSY degrees of freedom 
in penguin-like diagrams can also lead to operators with 
dimension lower than 6, creating an effective FCNC coupling 
between quarks and SM gauge fields. 
In the case of the photon field, 
the unbroken electromagnetic gauge 
invariance implies that the lowest-dimensional coupling 
is provided by the so-called magnetic operator
\beq
\bar{d}^i_{L(R)}\sigma^{\mu\nu} d^j_{R(L)} F_{\mu\nu}~.
\label{eq:mag}
\eeq
Here the potentially leading SUSY contribution is 
induced by helicity-violating couplings, and in particular 
by the left--right mixing of down-type squarks, which 
can appear in gluino-exchange diagrams:
\beq
\frac{ (\delta^D_{LR})_{ij}  }{ M_S}~.
\label{eq:LRm1}
\eeq
Since the operator (\ref{eq:mag}) has dimension 5, 
the explicit dimensional suppression of the 
left--right mixing contribution is only  
$1/M_S$. Nonetheless, also in this case the  
overall SUSY effect decouples as $1/M^2_S$ 
since  $\delta^Q_{LR}= \cO(m_q /M_S)$. 

The appearance of a single inverse power of $M_S$
in Eq.~(\ref{eq:LRm1}) has the important consequence
that this contribution can naturally evade the 
$\Delta F=2$ constraints and compete with the SM 
term \cite{GGMS,MM,BCIRS}. This is not the case for 
contributions generated 
by helicity-conserving couplings
or left--right mixing in the up sector, which appear 
only beyond the first order in the mass insertion.

In the $b\to s$ case the most significant constraint on 
possible non-standard effects in the magnetic operator 
is provided by the inclusive process $B\to X_s \gamma$
(see e.g. \cite{bsg} for an updated discussion).
The recent measurements \cite{bsgexp} exclude 
SUSY contributions substantially larger that the SM 
one, or imply bounds of $\cO(10^{-2})$  on $|(\delta_{LR}^D)_{23}|$.
Note, however, that the assumption made on the trilinear
terms implies 
\beq
\left|(\delta_{LR}^D)_{23} \right| \lsim { m_b \over M_S } \simeq
10^{-2} \left( \frac{500~{\rm GeV}}{M_S} \right)~,
\eeq
then the $B\to X_s \gamma$ measurement does 
not pose a serious fine-tuning constraint
about the non-universality of  $A$ terms.

Concerning the $s\to d$ case, there are no significant 
constraints on $|(\delta_{LR}^D)_{12}|$, whereas a 
stringent bound on  $|\Im(\delta_{LR}^D)_{12}|$ can be 
derived from $\varepsilon^\prime/\varepsilon$ \cite{MM}.
The latter is obtained by constraining  the SUSY contribution 
to the chromo-magnetic operator (closely related to the 
magnetic one) and implies~\cite{BCIRS}:
\beq
\left|\Im(\delta_{LR}^D)_{12} \right| \leq 4 \times 10^{-5}
\left( \frac{M_S}{500~{\rm GeV}} \right)~.
\eeq
This limit is more stringent than the 
upper bound on $|(\delta_{LR}^D)_{12}|$
imposed by (\ref{CDcond}), namely 
\beq
\left|(\delta_{LR}^D)_{23} \right| \lsim { m_s \over M_S } \simeq 2 \times
10^{-4} \left( \frac{500~{\rm GeV}}{M_S} \right)~,
\eeq
but it is much higher than the value 
assumed by $|\Im(\delta_{LR}^D)_{12}|$ within the 
flavour-constra\-ined MSSM \cite{Silve_new}.
Interestingly, if $\Im(\delta_{LR}^D)_{12}= 
\cO( 10^{-5})$ it is possible to 
conceive a scenario where all CP-violating effects
observed so far in the kaon sector 
($\varepsilon$ and $\varepsilon^\prime$)
are of SUSY origin \cite{mura2,DIM}.
As we shall discuss in the next sections, this
scenario would produce very clear signatures 
in rare  $K$ decays.

\item[$Z$ penguins.] 
Thanks to the spontaneous breaking of $SU(2)_L$, 
in the case of $Z$ penguins the integration of the SUSY degrees of 
freedom can lead to an effective FCNC operator of dimension 4: 
\beq
\bar{q}^i_L \gamma^\mu d^j_L Z_\mu~.
\label{eq:Zpeng}
\eeq
This operator generates a dimension-6 structure like 
the one in Eq.~(\ref{eq:dim6}) when the heavy $Z$ field 
is integrated out. In this case, however, the dimensional
suppression is provided by $1/M_Z^2$ and there is no 
explicit trace of $M_S$.  The latter is hidden in the 
dimensionless coupling of the operator (\ref{eq:Zpeng}),
denoted by $Z^L_{ji}$,
that requires a double mixing between  
$SU(2)_L$-singlet and $SU(2)_L$-doublet 
fields,\footnote{Here and in the following we employ 
the normalization of $Z^L_{ji}$ in \cite{BCIRS,BHI}:
$$
\cL^{Z}_{\rm FC} = \frac{G_F}{\sqrt{2}} \frac{e}{ \pi^2} M_Z^2
 \frac{\cos \Theta_W}{\sin \Theta_W} Z^\mu   
  Z^L_{ji}~\bar q^i_L \gamma_\mu q^j_L   \,+\, {\rm h.c.}~.
$$
With this normalization the SM contribution 
to $Z^L_{ji}$, evaluated in the  't~Hooft--Feynman gauge, 
is given by  $Z^L_{ji}\vert_{\rm SM} \simeq C_0(x_t) V_{3i}^* V_{3j}$,
where $V_{ij}$ denote CKM matrix elements, $x_t=m_t^2/m_W^2$ 
and the function $C_0(x)$ can be found in \cite{BBL}.
We further stress that the leading $\cO(x_t)$
contributions to FCNC $Z$ penguins are gauge-invariant within both 
SM and MSSM.}
and thus vanishes 
as $1/M^2_S$ for large $M_S$. The potentially
dominant contributions to $Z^L_{ji}$ arise 
from chargino loops, 
either by a double left-right insertion in the up-squark propagators
\cite{CI} or by a single insertion together with wino-higgsino 
mixing \cite{BRS,LMSS}:
\beq
Z^L_{ji} \sim \left\{ \ba{l} (\delta^U_{LR})_{j3}  (\delta^U_{RL})_{3i} \\
 (m_t/M_S) V_{j3} (\delta^U_{RL})_{3i} \ea \right.
\eeq
As can be noted, in both cases $Z^L_{ji} = \cO(m_t^2/M^2_S)$,
where the $m_t$ factor arises from the Yukawa coupling of  
the third generation.
Since the left--right mixing in the up sector provides a 
subleading contribution to generic dimension-6
operators and, in particular, to $\Delta F=2$
transitions, the indirect constraints on these effects 
are rather weak. If $(\delta^U_{RL})_{3i}$ lies in the window
\beq
\frac{m_t}{M_W} |V_{3i}| \lsim \left|(\delta^U_{LR})_{3i}\right| \lsim
\frac{m_t}{M_S}~,
\label{eq:window}
\eeq
then SUSY contributions to $Z^L_{ji}$ turn out to be comparable 
or even larger than the SM one. 
On the contrary, contributions to $Z^L_{ji}$ from 
helicity-conserving couplings or left--right mixing in the 
down sector are always negligible. 

In the $b\to s$ case some phenomenological
constraints on $|Z^L_{sb}|$ can be obtained directly
from exclusive and inclusive $b \to s~\ell^+\ell^- (\nu\bar\nu)$ 
transitions \cite{LMSS,BHI}. The latter are certainly cleaner from the 
theoretical point of view; however, their experimental determination 
is quite difficult. Indeed  the  most stringent 
constraint, at present, is the one extracted 
from $B\to K^* \mu^+\mu^-$ \cite{BHI}, 
where the  experimental upper bound on the non-resonant branching 
ratio lies only about 
a factor of 2 above the SM expectation. This constraint 
implies a bound of $\cO(1)$ on  $|(\delta^U_{LR})_{3i}|$,
which is still outside the window (\ref{eq:window}).

Owing to the smallness of $V_{td}$, 
the window (\ref{eq:window}) is much larger in the case of
$s\to d$ transitions. Here the most stringent constraints on 
$Z^L_{ds}$ arise from $K_L \to \mu^+ \mu^-$ (on the real part)
and $\varepsilon^\prime/\varepsilon$  (on the imaginary part)
\cite{BS}. Without entering into a detailed discussion about 
these bounds, which can be found elsewhere \cite{BCIRS}, we 
simply note that: i) the sizeable uncertainties due to  
non-perturbative effects in both $K_L \to \mu^+ \mu^-$ and 
$\varepsilon^\prime/\varepsilon$ do not allow us to extract precise 
constraints; ii)~taking into account these 
uncertainties, the present bounds on $Z^L_{ds}$ 
are within the 
window (\ref{eq:window}) and allow for $\cO(1)$ 
deviations from the SM at the amplitude level. 

\end{description}

\noindent
Summarizing, we can say that only the flavour-violating 
left--right mixing among the squarks can naturally 
 lead to large effects in the 
transitions (\ref{eq:uno}). In the $s\to d$
case this can happen either via magnetic penguins 
[ruled by $(\delta^D_{LR})_{12}$] or via $Z$ penguins  
[ruled by $(\delta^U_{LR})_{13}$ and $(\delta^U_{LR})_{23}$],
whereas  $b\to s$  magnetic penguins are 
strongly constrained by $B\to X_s \gamma$.
Moreover, we have seen that under the assumption
(\ref{CDcond}) the present constraints 
about the non-universality of the trilinear terms are all 
rather weak, both for up- and down-type squarks. 
We believe that this observation strengthens the interest 
in searching for sizeable non-standard effects in the 
transitions~(\ref{eq:uno}). 

\section{$K \to \pi \nu\bar\nu$}
These decays are considered the golden modes for a precise 
measurement of the $s \to d \nu \bar{\nu}$ transition.
Within the SM, separating the contributions to the 
 $s \to d \nu \bar{\nu}$  amplitude according to the 
intermediate up-type quark running inside the loop, one can write
\beq 
\cA(s \to d \nu \bar{\nu})_{\rm SM} = \sum_{q=u,c,t} V_{qs}^*V_{qd} \cA_q 
\sim \left\{ \begin{array}{ll} \cO(\lambda^5
m_t^2)+i\cO(\lambda^5 m_t^2)\    & (q=t) \\
\cO(\lambda m_c^2 )\ + i\cO(\lambda^5 m_c^2)     & (q=c) \\
\cO(\lambda \Lambda^2_{QCD})    & (q=u)
\end{array} \right. \!
\label{uno}
\eeq
The hierarchy of the CKM matrix elements
would favour  up- and charm-quark contributions;
however, the hard GIM mechanism of the parton-level calculation
implies $\cA_q \sim m^2_q/M_W^2$, leading to a completely 
different scenario. As shown on the r.h.s.~of Eq.~(\ref{uno}), 
where we have employed the standard phase convention 
($\Im V_{us}=\Im V_{ud}=0$) and 
expanded the CKM matrix in powers of the 
Cabibbo angle ($\lambda=0.22$) \cite{Wolf},
the top-quark contribution dominates both real and
imaginary parts.
This structure implies that $\cA(s \to d \nu \bar{\nu})_{\rm SM}$
is dominated by short-distance dynamics and therefore calculable 
with high precision in perturbation theory.

The leading short-distance contributions to $\cA(s \to d \nu \bar{\nu})$,
both within the SM and within its SUSY extension discussed before, 
can be described by means of a single effective dimension-6 operator:
\beq
Q^{\nu}_{L}= (\bar{s}_L\gamma^\mu d_L)(\bar{\nu}_L \gamma_\mu \nu_L)~,
\eeq 
whose Wilson coefficient has been calculated at the next-to-leading order 
within the SM \cite{BB} (see also \cite{MU,BB2}).
The simple structure of $Q^{\nu}_{L}$ has two major advantages: 
\begin{itemize}
\item{} the relation between partonic and hadronic amplitudes 
is very accurate, since the hadronic matrix elements
of the $\bar{s} \gamma^\mu d$ current between a kaon and a pion
are related by isospin symmetry to those entering $K_{l3}$ 
decays, which are experimentally well known; 
\item{} the lepton pair is produced in a state of definite CP 
and angular momentum, implying that the leading contribution 
to $K_L \to \pi^0  \nu \bar{\nu}$ is CP-violating.
\end{itemize}

The dominant theoretical error in estimating 
$\Br(K^+\to\pi^+ \nu\bar{\nu})_{\rm SM}$
is due to the uncertainty of the QCD
corrections to the charm contribution
(see \cite{BB2} for an updated discussion), which 
can be translated into a $5\%$ error in the determination
of $|V_{td}|$ from 
$\Br(K^+\to\pi^+ \nu\bar{\nu})$~\footnote{Very 
recently also the subleading effect of 
$\cO(m^2_K/m_c^2)$ induced by dimension-8 operators has 
been estimated \cite{Falk}. This effect is not calculable 
precisely, but it is likely to be smaller than 
(or at most as large as) the uncertainty 
in the QCD corrections to the leading term \cite{Falk}.}.
Genuine long-distance effects associated to the up quark 
have been shown to be substantially smaller
\cite{LW}.

The case of $K_L\to\pi^0 \nu\bar{\nu}$ is even cleaner from the
theoretical point of view \cite{Litt}. Indeed, because of the  CP
structure, only the imaginary parts in (\ref{uno}) 
--where the charm contribution is absolutely negligible--
contribute to $\cA(K_2 \to\pi^0 \nu\bar{\nu})_{\rm SM}$. Thus 
the dominant direct-CP-violating component 
of $\cA(K_L \to\pi^0 \nu\bar{\nu})_{\rm SM}$ is completely saturated by 
the top contribution, where the QCD uncertainties are 
very small (around 1\%). 
Intermediate and long-distance effects in this process
are confined to the indirect-CP-violating 
contribution \cite{BB3} and to the CP-conserving one 
\cite{CPC} which are both extremely small.
Taking into account also the isospin-breaking corrections to the hadronic
matrix element \cite{MP}, one can
write an expression for $\Br(K_L\to\pi^0 \nu\bar{\nu})_{\rm SM}$ 
in terms of short-distance parameters with a theoretical error below $3\%$
\cite{BB2,BB3}: \beq
\Br(K_L\to\pi^0 \nu\bar{\nu})_{\rm SM}~=~4.16 \times 10^{-10}~\left[
\frac{\overline{m}_t(m_t) }{ 167~{\rm GeV}} \right]^{2.3} ~\left[ \frac{\Im
\lambda_t }{ \lambda^5 } \right]^2~, \eeq
where $\lambda_t = V_{ts}^*V_{td}$.

Taking into account all the indirect constraints on 
$\Im(V_{ts}^*V_{td})$ \cite{Dago}, 
the present range of SM predictions for the two 
$K\to \pi \nu\bar \nu$ branching ratios is reported in the second 
column of Table~\ref{tab:rare}. In the following three columns, we 
show the upper bounds obtained within three SUSY 
scenarios with non-trivial   $(\delta^U_{LR})_{i3}$
and $(\delta^U_{LR})_{12}$.
In all cases the SUSY flavour-mixing terms, as well as 
CKM matrix elements, have been constrained
taking into account the measurement of $\varepsilon$,
$\varepsilon^\prime$, $K_L \to \mu^+\mu^-$ and the 
respective theoretical uncertainties \cite{BCIRS}.
As can be noticed, the two neutrino modes
could provide sizeable unambiguous signatures of SUSY, 
but only in the presence of a large left--right mixing in 
the up sector. Interestingly, the 
present measurement of 
$\Br(K^+\to \pi^+  \nu\bar \nu)$ \cite{nt}
is very close to putting serious 
constraints (or to providing some evidence\ldots) 
on this scenario.

\begin{table}[t]
\begin{center}
  \begin{tabular}{|l|l|l|l|l|l|}  \hline   \multicolumn{1}{|c|}{Observable} 
  & \multicolumn{1}{c|}{SM} & \multicolumn{3}{c|}{SUSY scenarios} 
  & \multicolumn{1}{c|}{exp. data} \\ 
  & &  \multicolumn{1}{c}{A} & \multicolumn{1}{c}{B} & \multicolumn{1}{c|}{C}  
  &  \\   \hline 
  $10^{10}\times\Br(\kppn) $   &  $0.71\pm 0.12$ & $\leq \Br_{\rm SM}$  & 
        $\leq 2.1$  &  $\leq 2.7$  &  $\quad  1.5^{+3.5}_{-1.3} $ 
  \protect\cite{nt} \\ 
  $10^{10}\times\Br(\klpn)$    &  $0.22\pm 0.05$ & $\leq \Br_{\rm SM}$  & 
        $\leq 1.7$  &  $\leq 4.0$  &  $<5.9\times 10^3$ 
  \protect\cite{KTeV0nn} \\ 
  $10^{11}\times\Br(\kpe)_{\rm dir}$   &  $0.35\pm 0.07$ & $\leq 2.0$  & 
        $\leq 3.0$  &  $\leq 10$  &  $< 58$ \protect\cite{KTeV0ee} \\ \hline
  \end{tabular}
\end{center}
  \caption{SM expectations, experimental data and upper bounds 
    within different 
    SUSY scenarios for the branching ratios 
    of the rare decays $\klpn$, $\kpe$ and $\kppn$. 
    The three SUSY scenarios correspond to \cite{BCIRS}: 
    A) $(\delta^U_{LR})_{i3}=0$, $(\delta^U_{LR})_{12}\not=0$, 
       ${0\leq \Im(\lambda_t) \leq  \Im(\lambda_t)_{\rm SM}}$;    
    B) $(\delta^U_{LR})_{12}=0$, $(\delta^U_{LR})_{i3}\not=0$,
       $0\leq \Im(\lambda_t) \leq  \Im(\lambda_t)_{\rm SM}$;     
    C)~${(\delta^U_{LR})_{12}\not=0}$, $(\delta^U_{LR})_{i3}\not=0$,
       $| \Im(\lambda_t) | \leq 1.73 \times 10^{-4}$.   
    \label{tab:rare} }
\end{table}

\section{$K_L \to \pi^0 e^+ e^-$}
Similarly to $K\to\pi\nu\bar{\nu}$ decays, 
also the short-distance contributions to 
$K\to\pi \ell^+\ell^-$ transitions are calculable with high accuracy.
Long-distance contributions to the latter, however, are much larger 
owing to the presence of electromagnetic interactions. 
Only in few cases (mainly in CP-violating observables)
are long-distance contributions suppressed and is it 
possible to extract the interesting short-distance
information. 

The single-photon exchange amplitude, dominated by long-distance dynamics,
provides the largest contribution to the CP-allowed transitions 
$K^+ \to \pi^+ \ell^+ \ell^-$ and $K_S \to \pi^0 \ell^+ \ell^-$.
The former has been observed, both in the electron and in the muon 
mode, whereas only an upper bound of $1.6 \times 10^{-7}$
exists on $\Br(K_S\to\pi^0 e^+e^-)$ \cite{Na48}.
On the contrary, the long-distance part of the
single-photon exchange amplitude is forbidden by CP invariance
in the $K_L \to \pi^0 \ell^+ \ell^-$ channels, which are 
much more interesting from the short-distance point 
of view (especially the electron mode).

In $K_L \to \pi^0 e^+ e^-$ we can distinguish three 
independent (and comparable) contributions: direct-CP-violating, 
indirect-CP-violating and CP-conserving.

The direct-CP-violating part of the $K_L \to \pi^0 e^+ e^-$ 
amplitude is very similar to the $K_L \to \pi^0 \nu \bar{\nu}$ one,
but for the fact that it receives an additional short-distance contribution 
from the photon penguin. Within the SM, this theoretically clean part 
of the amplitude leads to \cite{BLMM}
\beq
\Br(K_L\to\pi^0 e^+e^-)^{\rm SM}_{\rm dir}~=~6.5 \times 10^{-11}~\left[
\frac{\overline{m}_t(m_t) }{ 167~{\rm GeV}} \right]^{2} ~\left[ \frac{\Im
\lambda_t }{ \lambda^5 } \right]^2~.
\eeq
The present range of variation, together with SUSY upper  
bounds, is reported in the last line of Table~\ref{tab:rare}.
Being sensitive also to the photon penguin, the 
$K_L \to \pi^0 e^+ e^-$ amplitude could be substantially 
modified also in the presence of non-trivial SUSY phases in the 
down sector. In particular, within the interesting scenario 
where all CP-violating effects observed in the kaon sector were 
due to  $\Im(\delta_{LR}^D)_{12} = \cO( 10^{-5})$,
$\Br(K_L\to\pi^0 e^+e^-)^{\rm SM}_{\rm dir}$
would be close to its SM value, whereas 
$\Br(K_L\to\pi^0 \nu\bar\nu)$ would be vanishingly small.

In principle the direct-CP-violating part 
of the $K_L \to \pi^0 e^+ e^-$ amplitude 
could be experimentally isolated from the other two 
contributions, especially if it were large.
In order to achieve this goal it would be necessary 
to measure $\Br(K_S \to \pi^0 e^+e^-)$
or to put a stringent bound on it.  
The two CP-violating 
components of the $K_L\to\pi^0 e^+e^-$ amplitude
will in general interfere, and the 
indirect-CP-violating one alone would lead to 
\beq
\Br(K_L \to \pi^0 e^+ e^-)_{\rm CPV-ind}~=~ 3\times 10^{-3}~ \Br(K_S \to \pi^0 e^+
e^-)~.
\eeq
Since the relative phase of the two  CP-violating 
amplitudes is known, once $\Br(K_S \to \pi^0 e^+ e^-)$ will be 
measured, it will be possible to determine the interference between direct
and indirect CP-violating components of $\Br(K_L\to\pi^0 e^+e^-)_{\rm
CP}$ up to a sign ambiguity. 

The  CP-conserving contribution 
to $K_L \to \pi^0 e^+ e^-$, 
generated by a two-photon intermediate state, 
does not interfere with the CP-violating ones  
and is expected to be in the $10^{-12}$ range. The relative weight of this 
contribution can be further constrained by appropriate kinematical cuts;
it should therefore not represent a problem if 
$\Br(K_L \to \pi^0 e^+ e^-)$ will be found 
in the $10^{-11}$ range.

\section{Exclusive $b\to s\ell^+\ell^-(\nu\bar{\nu})$ decays}
The starting point for the analysis of $b \to s \ell^+\ell^-(\nu\bar{\nu})$ 
transitions, both within the SM and the SUSY scenario discussed in Section~2, 
is the following effective Hamiltonian:
\beq
{\cal{H}}_{\rm eff} = - \frac{G_F}{\sqrt{2}}  V_{t s}^\ast  V_{tb}  
\left( \sum_{i=1}^{10} \left[ C_i  \op_i +
C_i^{\prime} \op_i^{\prime} \right] + C_L^\nu \op_L^\nu + C_R^\nu \op_R^\nu 
\right) \,+\, {\rm h.c.}~.
\label{eq:he}
\eeq
Here $\op_i$ denotes the Standard Model basis of operators relevant 
to $b \to s \ell^+ \ell^-$ \cite{BBL} and $\op_i^\prime$ their helicity 
flipped counterparts. In particular, we recall that 
$\op_i\sim(\bar s\gamma_\mu b)(\bar c \gamma^\mu c)$, for $i=1 \ldots 6$, 
$Q_7\sim m_b \bar s_L (\sigma\cdot F) b_R$, 
$Q_8\sim m_b \bar s_L (\sigma\cdot G) b_R$, 
$Q_9\sim (\bar s_L \gamma_\mu b_L)(\bar \ell \gamma^\mu \ell)$, 
$Q_{10}\sim (\bar s_L \gamma_\mu b_L)(\bar \ell \gamma^\mu \gamma_5 \ell)$
and  $Q_{L(R)}^\nu \sim (\bar s_{L(R)} \gamma_\mu b_{L(R)})(\bar \nu_L \gamma^\mu \nu_L)$.
The operators that have a non-vanishing  matrix element already at the tree level
and thus play the dominant role in $b \to s \ell^+ \ell^-$
are $Q_7$, $Q_9$, $Q_{10}$ and their helicity flipped counterparts.
On the other hand,  only $ Q_{L(R)}^\nu$ have a non-vanishing  matrix element 
in $b \to s \nu \bar \nu$.

Rate and CP asymmetry in $B\to X_s \gamma$ already provide 
serious constraints on possible deviations from the SM 
in $C_7$ and $C_7^\prime$ \cite{bsg},
and these bounds will soon improve with new data on $B\to X_s \gamma$.
However, as we have discussed in Section~2, even if no new-physics 
effects are found in the magnetic operator, one could still 
expect sizeable SUSY contributions mediated by the $Z$ penguin. 
In the following we shall concentrate only on the latter 
type of effects. Under this assumption, a rather simplified 
scenario emerges, where $C_R^\nu=C_i^\prime=0$ and 
only $C_{10}$ and $C^\nu_L$ are 
substantially modified from their SM value \cite{BHI}.

Moreover, even though inclusive measurements are certainly 
more suitable for precise determinations of  
short-distance parameters, here we shall 
discuss only exclusive decays, which have 
a clear  advantage from the experimental point of view. 
Within the SM the following exclusive branching ratios
are expected, compared here with the current
experimental limits:
\begin{equation}
\ba{rclll}
\Br(B\to K\nu\bar\nu)&\approx& 4\times 10^{-6}\qquad 
 &(< 7.7\times 10^{-4} &\cite{delphi96}) \\
\Br(B\to K^*\nu\bar\nu)&\approx& 1.3\times 10^{-5}\qquad 
 &(< 7.7\times 10^{-4} &\cite{delphi96})\\
\Br(B\to K\mu^+\mu^-)^{\rm n.r.}&\approx& 6\times 10^{-7} 
 &(< 5.2 \times 10^{-6}  &\cite{CDF})\\
\Br(B\to K^*\mu^+\mu^-)^{\rm n.r.}&\approx& 2\times 10^{-6}
 &(< 4\times 10^{-6}   &\cite{CDF})\\
\Br(B_s\to\mu^+\mu^-)&\approx& 3\times 10^{-9} 
 &(< 2.6\times 10^{-6} &\cite{CDF98})
\ea
\end{equation}
The corresponding hadronic uncertainties are typically around 
$\pm 30\%$ (see e.g. \cite{ABHH99} for an updated discussion).
As already mentioned, the channel that sets the strongest 
constraint on the FCNC $Z$ penguin is $B\to K^*\mu^+\mu^-$.
In the optimistic case where $Z^L_{bs}$
is close to saturating this bound, we 
would be able to detect the presence of non-standard
dynamics already by observing sizeable rate enhancements 
in the above listed branching ratios. In processes such as 
$B\to K^* \ell^+ \ell^-$ and $B\to K \ell^+ \ell^-$,
where the standard photon-penguin diagrams provide a 
large contribution, the enhancement could be at most a factor of 2. 
On the other hand, in processes such as
$B\to K^* \nu \bar{\nu}$,  $B\to K \nu \bar{\nu}$
and $B_s \to \mu^+ \mu^-$, where the photon-exchange 
amplitude is forbidden, the maximal enhancement could reach a 
factor of 10 \cite{BHI}. 

\subsection{Forward-backward asymmetry in $B\to K^*\mu^+\mu^-$}
If SUSY effects were not large enough to produce sizeable 
deviations in the magnitude of the  
$b \to Z^* s$ transition, as expected unless 
 $|(\delta^U_{LR})_{32}|$ were very close to 
the upper bound in Eq.~(\ref{eq:window}), 
it would be hard to 
detect them from exclusive rate measurements.
A more interesting observable in this respect 
is provided by the forward--backward (FB) asymmetry of the
emitted leptons.
In the  $\bar B\to \bar K^* \mu^+\mu^-$ case this is defined as 
\beqa
&& \cA^{(\bar B)}_{\rm FB}(s)=\frac{1}{d\Gamma(\bar B\to \bar K^* \mu^+\mu^- )/ds}
  \int_{-1}^1 \!\!\! d\cos\theta 
\no\\
&&  
\label{eq:asdef} \qquad
\frac{d^2 \Gamma(\bar B\to \bar K^* \mu^+\mu^- )}{d s~ d\cos\theta}
\mbox{sgn}(\cos\theta)~,
\eeqa
where $ s =m_{\mu^+\mu^-}^2/m_B^2$ and 
$\theta$ is the angle between the momenta of 
$\mu^+$ and $\bar B$ in the dilepton centre-of-mass frame. 
Assuming that the leptonic current has only a 
vector ($V$) or axial-vector ($A$) structure, then the
FB asymmetry provides a direct measure of 
the $A$-$V$ interference. 
Since the vector current is largely dominated by 
the photon-exchange amplitude and the axial one is 
very sensitive to the $Z$ exchange, 
$\cA^{(\bar B)}_{\rm FB}$ and $\cA^{(B)}_{\rm FB}$
provide an excellent tool
to probe the  $Z\bar{b}s$ vertex.
Indeed 
$\cA^{(\bar B)}_{\rm FB}(s)$ turns out to be 
proportional to\footnote{To simplify the notation we have 
introduced the parameter $\cne(s)$, which is not a Wilson 
coefficient but can be identified with $C_9$ at the 
leading-log level \cite{BHI}.}
\beq
  {\rm Re}\left\{  \ct^* \left[ s~\cne(s) 
    + \alpha_+(s) \frac{m_b C_7}{m_B}  \right] \right\}, 
  \label{eq:dfbabvllex}
\eeq
where $\alpha_+(s)$ is an appropriate ratio 
of hadronic form factors \cite{BHI,burdman0}.
The overall factor ruling the magnitude of $\cA^{(\bar B)}_{\rm FB}(s)$
is affected by sizeable theoretical 
uncertainties. Nonetheless there are at least 
three features of this observable
that provide a clear short-distance information:

i) Within the SM $\cA^{(\bar B)}_{\rm FB}(s)$ has a zero
in the  low-$s$ region ($s_0|_{\rm SM} \sim 0.1$) \cite{burdman0}.
The exact position of $s_0$ is not free from 
hadronic uncertainties at the $10\%$ level; 
nonetheless, the existence of the zero itself is 
a clear test of the relative sign between 
$C_7$ and $C_9$. The position of $s_0$ is 
essentially unaffected by possible
new-physics effects in the  $Z\bar{b}s$ vertex.

\begin{figure}
\centerline{\epsfysize=2.5in{\epsffile{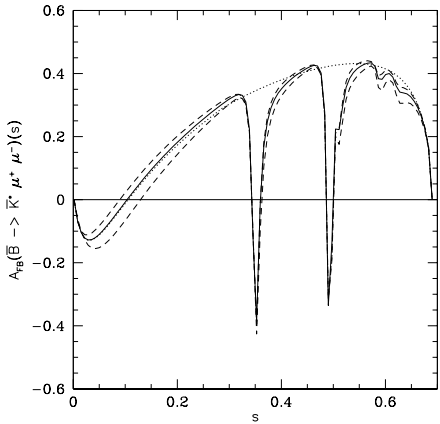}}}
\caption{ \it FB asymmetry of
$\bar B\to \bar K^* \mu^+\mu^-$ within the SM.
The solid (dotted) curves have been
obtained employing the Krueger--Sehgal \protect\cite{ks96} approach 
(using the perturbative end-point effective 
Hamiltonian \protect\cite{BHI}). 
The dashed lines show the effect of varying the
renormalization scale of the Wilson coefficients
between $m_b/2$ and $2 m_b$,
within the Krueger--Sehgal approach. }
\label{fig:AFB}
\end{figure}

ii) The sign of $\cA^{(\bar B)}_{\rm FB}(s)$ around the zero
is fixed unambiguously in terms of the relative sign
of $C_{10}$ and $C_9$ \cite{BHI}: within the SM one 
expects $\cA^{(\bar B)}_{\rm FB}(s) > 0$ for $s>s_0$,
as in Fig.~\ref{fig:AFB}.
This prediction is based on a model-independent 
relation between the form factors \cite{LEET}.
Interestingly, the sign of $C_{10}$
could change in the presence of a non-standard   
$Z\bar{b}s$ vertex, leading to a striking signal
of new physics in $\cA^{(\bar B)}_{\rm FB}(s)$, 
even if the rate of $\bar B \to \bar K^* \mu^+\mu^-$
were close to its SM value.

iii) In the limit of CP conservation, one expects 
$\cA^{(\bar B)}_{\rm FB}(s) = - \cA^{(B)}_{\rm FB}(s)$.
This holds at the per-mille level within the 
SM, where $C_{10}$ has a negligible CP-violating phase,
but again it could be different in the presence 
of new physics in the  $Z\bar{b}s$ vertex.
In this case
the ratio $[ \cA^{(\bar B)}_{\rm FB}(s) + \cA^{(B)}_{\rm FB}(s)]/
[ \cA^{(\bar B)}_{\rm FB}(s) - \cA^{(B)}_{\rm FB}(s)]$
could be different from zero, for $s$ above the charm threshold, 
even reaching the $10\%$ level in
the SUSY scenario of Section 2 \cite{BHI}. 

\section{Conclusions}
Rare FCNC transitions of the type $d_j \to d_i~\ell^+\ell^-(\nu\bar\nu)$ 
are very sensitive to simultaneous violations of $SU(2)_L$ and 
flavour symmetries. Within generic supersymmetric extensions 
of the SM, these processes could be substantially modified 
 in the presence of non-diagonal trilinear 
soft-breaking terms.
At present this possibility is still open 
for both $b \to s$ and $s\to d$ transitions, but it has 
more chances to be realized in the $s\to d$ case \cite{Vives}.
The future measurements of $\Br(\kppn)$,  
$\Br(\klpn)$, $\Br(\kpe)$ and 
$\cA_{\rm FB}[B(\bar B) \to \bar K^* \mu^+\mu^-]$
will provide very useful insights
in this scenario.

\section*{Acknowledgements}
It is a pleasure to thank the organisers of RADCOR 2000
for their invitation to this interesting symposium
and their financial support.

\end{document}